\documentclass{article}
\usepackage{graphicx} 
\usepackage{amsmath,amssymb,amsfonts,amsthm}
\usepackage{float}
\usepackage[a4paper, total={7in, 10in}]{geometry}
\usepackage{subcaption}
\usepackage[sorting=none]{biblatex}
\usepackage[hidelinks]{hyperref}
\title{\textbf{Alternative threshold function for Bayesian Optimization of Variational Quantum Circuits}}
\author{Shreyas Dillon}
\date{May 2025}
\begin{document}
\maketitle
In this paper, we propose an expansion of the \textit{Expected Maximum Improvement over Confident Regions} (EMICoRe) Variational Quantum Eigensolver (VQE)\textemdash a technique advanced by Nicoli \emph{et. al.} \cite{emicore} which utilizes both quantum and classical components to approximate the ground state of a quantum system\textemdash by introducing an alternative threshold for EMICoRe’s Confident Region that depends on both the Gaussian process (GP) prior variance and the model's change in predicted energy over a set number of iterations. This modification is a more lenient threshold for the Confident Region and accounts for natural fluctuations in the predicted energy that EMICoRe punishes by eliminating the exploratory benefits presented by the Confident Region. We test both algorithms with the original EMICoRe model as a baseline and our results suggest improvement over EMICoRe's state-of-the-art results for a common benchmark for VQE's, the Ising Hamiltonian, and similar performance for more complex optimization regimes. We analyze the accuracy in approximated ground state energy and how the threshold evolves during optimization to compare the EMICoRe model with the proposed alternative. After comparison, we discuss the potential optimization of the degrees of freedom present in the new threshold for better performance and a more varied choice of system to be approximated.
\section{Introduction}
 \indent In recent times, there has been an immense push forward for quantum technology, quantum hardware, and specifically quantum algorithms \cite{computing}\cite{qualgo}\cite{qutech}. These algorithms have the potential to decrease the computation time necessary for a diverse set of applications, including cryptography (Shor’s algorithm) \cite{shor} and quantum chemistry \cite{cao}.  However,  with quantum computing currently in the Noisy Intermediate-Scale Quantum (NISQ) era \cite{nisq} we are not yet able to fully rely on quantum technology to be an improvement over current classical methods \cite{challengenisq}. NISQ devices are limited in the amount of qubits they are able to maintain, are not advanced enough to achieve fault tolerance or error correction, and are characterized by their susceptibility to quantum decoherence \cite{qd}. Due to this vulnerability and the difficult nature of controlling the states of many qubits \cite{challenges}\cite{google}, the width and depth of implementable quantum circuits are limited and thus hybrid quantum-classical algorithms have been employed to leverage quantum resources while minimizing the quantum degrees of freedom~\cite{hybrid}.

\section{About VQEs and Bayesian Optimization}
\subsection{Variational Quantum Eigensolvers}
\indent VQEs are one approach to numerically calculating the ground state energy and wavefunction for a given quantum Hamiltonian with $N$ qubits \cite{vqe}. Similarly to how a neural network models functions through parametric layers, VQEs use parametric quantum circuits to model wave functions. Essentially, the quantum computer prepares a trial state $\psi_t$ by acting on an initial state $\psi_0$ with $N_g$ unitary quantum gates dependent on angular parameters \cite{circuits}. $N_p$ of these quantum gates are parametrized with one of the $N_p$ angular parameters exclusively where each parameter $\alpha_d\in [0,2\pi)$ and $N_g \geq N_p$. The quantum computer then calculates the expectation value of the Hamiltonian for this state and the classical computer uses this estimate of the output to optimize the system to find the set of parameters for which the expectation value of the Hamiltonian (average energy of the state) is minimized, see Figure \ref{fig:15} and~\cite{emicore}. \\
\subsection{Bayesian Optimization}
\indent In order to optimize the parameters of the VQE to best approximate the ground state, numerous techniques utilizing Bayesian Optimization (BO) have been popularized, differing most notably in the choice of surrogate and acquisition functions \cite{frazier,acquisition}. The model we are looking to improve introduces the acquisition function EMICoRe (Expected Maximum Improvement over Confident Regions). In traditional BO, a surrogate function (most commonly Gaussian process (GP) regression \cite{gaussianprocess}) is used to approximate the true system, and then a separately chosen acquisition function attempts to find the most promising points using the current version of the surrogate function. Those points deemed most promising to the optimization are observed, incorporated into the GP's training set of points, and then the cycle repeats with the acquisition function finding the best points from the now updated GP. Using this approximation function and the best points yielded by BO, we can optimize the posterior GP after each iteration with the ultimate goal of finding the angular parameters that most closely approximate the ground state of the system. For more details on how this optimization can be conducted and how this method works, see Figure \ref{fig:15} and~\cite{emicore}.
\begin{figure}[H]
    \centering
    \begin{subfigure}[b]{0.45\textwidth}
    \includegraphics[width=\linewidth]{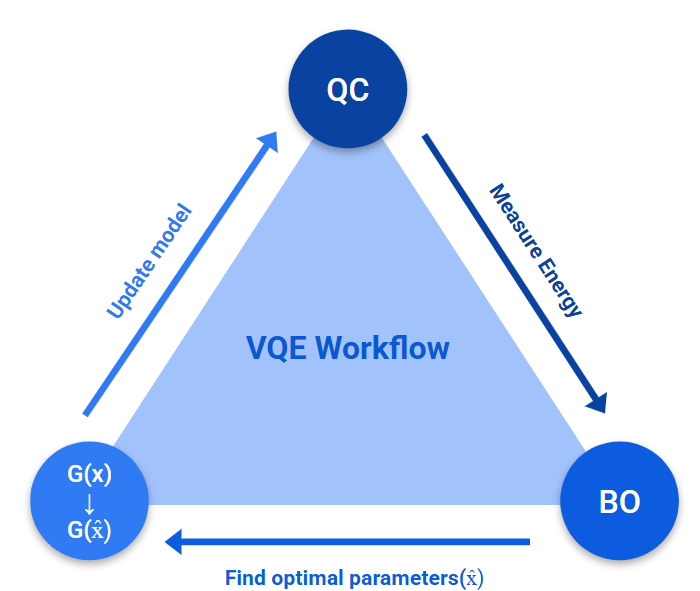}
    \caption{}
    \label{fig:15}
    \end{subfigure}
    \hfill
    \begin{subfigure}[b]{0.45\textwidth}
    \centering
    \includegraphics[width=\linewidth]{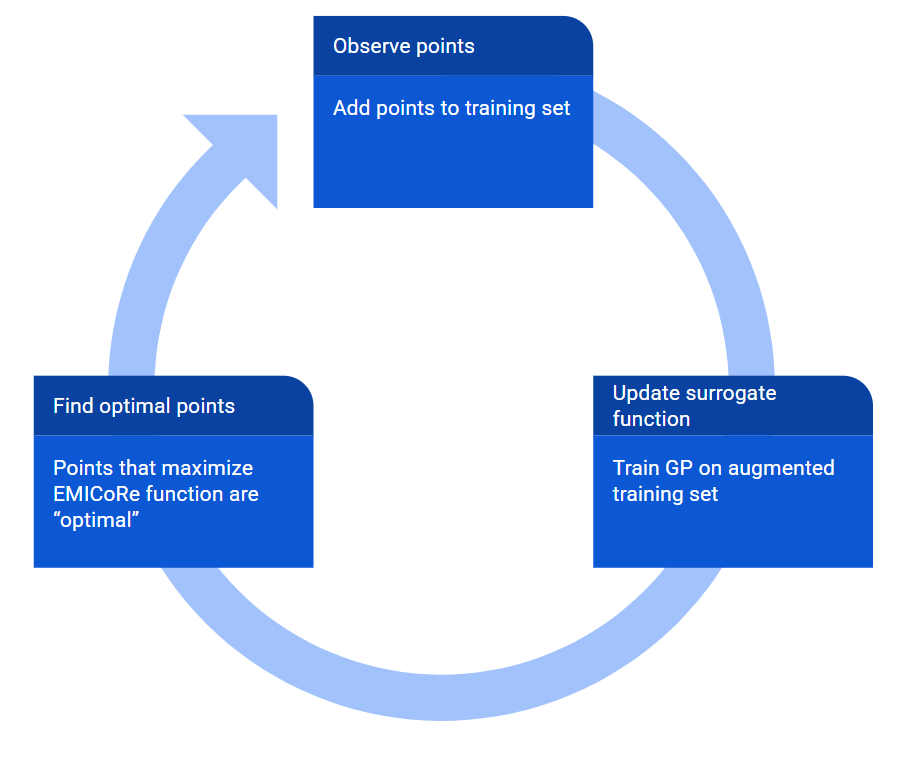}
    \caption{}
    \label{Figure 15}
    \end{subfigure}
    \caption{The VQE workflow (a) and BO workflow (b). For the VQE workflow, every iteration the quantum computer estimates the energy of the approximated system $G(x)$. Then through BO the optimal parameters $\hat{x}$ are found and the system is updated: $G(x)\rightarrow G(\hat{x})$. The energy of this new system is estimated and the next round begins. 
    The BO workflow is emphasized in (b) to highlight the steps that go into finding the optimal parameters $\hat{x}$}.
    \label{fig:15}
\end{figure}
\section{The EMICoRe Model}
\indent Prior to the EMICoRe model, the optimum from the previous iteration and other deterministically or (quasi) randomly chosen points were used for the next iteration \cite{nft} \cite{random}. Although this model was effective, the deterministic approach did not allow for any optimization of the points used in the following iterations, leading to a slower convergence rate to the ground state and less accurate approximations of both the ground state wave function and the ground state energy over a set number of iterations \cite{emicore}.

\indent One way the EMICoRe model changed this was by introducing the Confident Region (CR). A significant advantage garnered by GPs is that the posterior comes with a predicted covariance for the outputs that does not depend on the outputs themselves. This means that, given a set of training points and the current posterior GP, you can quantify the uncertainty of the true values of new points before observing them (for more information on GPs and this advantage, see \cite{gaussianprocess}).  In the EMICoRe model, if a point has a predicted variance below a chosen threshold value $\kappa^2$,  it can be treated as indirectly observed and added to a set called the Confident Region. Each iteration, every point in this region is tested by estimating the output of the acquisition function, and the points that are deemed the most promising (the points in the Confident Region that maximize the acquisition function) are included in the next round of optimization; see \cite{emicore}.\\

\indent In Ref. \cite{emicore}, the root of the threshold, $\kappa$, was set to be the energy decrease over a chosen number of iterations ($T_{Avg}$), given by 
$$\kappa^{t+1}=\frac{\mu^{t-T_{Avg}}-\mu^t}{T_{Avg}}$$ where $\mu^t=\mu(\mathbf{x}^t)$ is the mean of the GP posterior at optimization iteration $t$, $\mu^{t-T_{Avg}}$ is the mean of the GP posterior at iteration $t-T_{Avg}$, and $\mathbf{x}^t$ is the current vector of angular parameters. Thus, we are taking the best score (the lowest energy), at the current iteration, subtracting the best score from $T_{Avg}$ iterations ago and dividing by $T_{Avg}$ to obtain the average decrease in energy over the $T_{Avg}$ iterations. This becomes the new threshold, and the optimization continues. The authors justified this choice by highlighting their intuition that optimization should be crude at first (a large $\kappa$), allowing a significant number of points to be considered, but as the model becomes more refined (a smaller $\kappa$), the allowed uncertainty for the CR should decrease correspondingly, only allowing points with very low predicted variance to be considered. This matches the trend seen in the change in energy values predicted by the model; see Fig. \ref{fig:Figure 3}. There are initially large decreases in energy followed by smaller oscillations as the model gets closer to reaching the true energy. Although this choice of $\kappa$ proved to be effective and the EMICoRe model demonstrated growth over the previous state-of-the-art model \cite{emicore}\cite{nft}, the choice of the threshold is based on a relatively trivial observation and led to a largely heuristically based threshold that is not representative of the importance of the threshold function.
\section{Prior and Energy Dependent Threshold (PEDT)}

\indent All tests were run using the same parameters for the algorithm as in the original EMICoRe paper, with the exception of the number of qubits comprising the system and the number of iterations for optimization. The system to be approximated was the 10 qubit Ising Hamiltonian, both off criticality and at criticality in the thermodynamic limit \cite{ising}\cite{emicore}. The Ising Hamiltonian at criticality is more difficult to optimize compared to off-criticality due to high levels of entanglement in the system, so it is included in the analysis to gauge performance in a more challenging regime. We compare our estimated ground state energy with the analytically computed values for these systems. For more about the exact parameters used see \cite{emicore} and \cite{emicorecode}.\\

\indent The main intuition driving the proposed threshold function was that in the EMICoRe function the threshold controlling the allowed variance was not related to a variance. Although using the energy decrease as a bound is effective in the middle stages of optimization, where the decrease is small enough to be practical for a variance value, in the beginning stages, especially for larger systems, the decrease in energy is particularly volatile. This volatility leads to more unreliable optimization as the threshold will be too large to be effective in the initial iterations, see Appendix A and Fig. \ref{fig:Figure 2}. Secondly, in the code attached to their paper, Nicoli \emph{et. al.} defined $\kappa=$max$(0,\frac{\mu^{t-T_{Avg}}-\mu^t}{T_{Avg}})$ \cite{emicorecode}. Ideally, the average energy would decrease and $\frac{\mu^{t-T_{Avg}}-\mu^t}{T_{Avg}}$ would be positive as that would represent a decrease in energy over $T_{Avg}$ but if the model increases in energy, i.e., gets further from the ground state, then the average energy decrease would be negative. With $\kappa\geq 0$ as it was in \cite{emicore}, any such negative values are eliminated, preventing them from affecting optimization in later iterations. However, both initially and in the final phases of optimization, it is natural for the model to return positive changes in energy, initially due to large fluctuations in energy and in the closing stages due to all energy values being very similar as the model converges; see Fig. \ref{fig:Figure 1}. However, when a positive energy change occurs and $\kappa$ is set to 0 in the EMICoRe model, the benefit gained through the CR is lost and convergence to the ground state is slower and less effective; see Fig. \ref{fig:Figure 1}.\\

\indent To attempt to solve both of these issues, a Prior and Energy Dependent Threshold (PEDT) is proposed and tested against EMICoRe. Under the standard GP regression assumptions adopted by the EMICoRe paper, the prior variance $\sigma_0^2$ can act as an upper bound for the posterior variance; see \cite{gaussianprocess} and Appendix B. Thus, we make $\kappa_{PEDT}\propto\sigma_0$ to ensure that $\kappa_{PEDT}$ is always on the same order as the variance of the chosen system. $\kappa_{PEDT}$ is not proportional to the variance ($\sigma_0^2$) as $\kappa^2$ not $\kappa$ is used in the EMICoRe model. More specifically, with $\delta_E=|\frac{\mu^{t-T_{Avg}}-\mu^t}{T_{Avg}}|$, we set \begin{equation}\label{eq:1}
        \kappa_{PEDT}=\frac{\sigma_0}{2}\cdot \text{min}(\delta_E,\frac{1}{3+e^{-\delta_E}}). \end{equation}
By utilizing the absolute value of the average energy decrease over $T_{Avg}$ we eliminate the instances where $\kappa$ was 0 in the EMICoRe model, instead using a more lenient threshold to account for natural fluctuations in the model to retain the advantages of the CR. We take the minimum of $\delta_E$ and $\frac{1}{3+e^{-\delta_E}}$ due to the behavior of $\frac{1}{3+e^{-\delta_E}}$ as it approaches 0, see Fig. \ref{fig:Figure 2}.\\
\begin{figure}[H]
    \centering
    \includegraphics[width=0.5\linewidth]{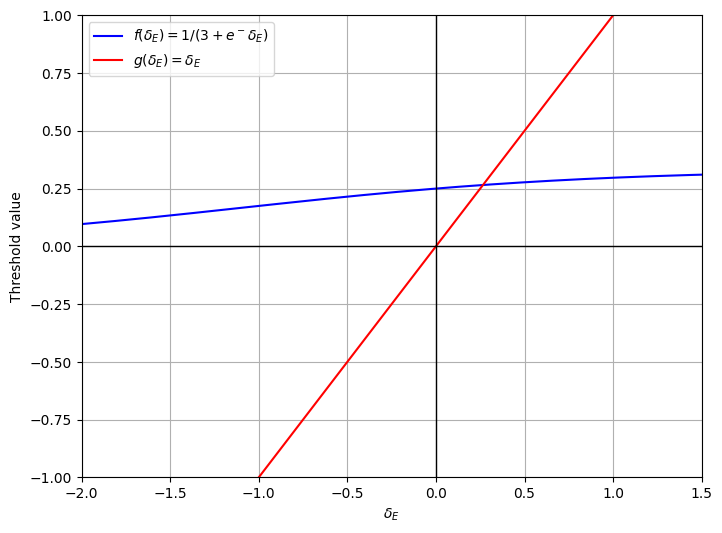}
    \caption{Graph of $f(\delta_E)=\frac{1}{3+e^{-\delta_E}}$ vs. $g(\delta_E)=\delta_E$. Near the origin $\frac{1}{3+e^{-\delta_E}}$ doesn't decrease fast enough for refined optimization but it is less than $g(\delta_E)$ for greater values of $\delta_E$ so initial optimization is ensured to be controlled.}
    \label{fig:Figure 2}
\end{figure}
\indent Fig. \ref{fig:Figure 2} highlights the motivation for two main parts of $\kappa_{PEDT}$. The reason behind obtaining the minimum of $\delta_E$ and $\frac{1}{3+e^{-\delta_E}}$ is that near the origin $\frac{1}{3+e^{-\delta_E}}$ decreases only slightly, which could hinder optimization in the later phases by allowing points with larger variance than desired into the CR. An additional feature of Fig. \ref{fig:Figure 2} is that for larger values of $\delta_E$, $g(\delta_E)=\delta_E$ increases significantly more than $\frac{1}{3+e^{-\delta_E}}$. Because we are altering the model to include multiplication by $\frac{\sigma_0}{2}$, which for a weak prior and any moderately sized system is likely to be greater than one (Nicoli \emph{et. al.} chose $\sigma_0$ to be on the same order as the number of qubits), we need a way to ensure that the threshold does not become too large, which would also ultimately lead to slower optimization, see Appendix A. Forcing the model to take the minimum of the two functions achieves this goal as $\frac{1}{3+e^{-\delta_E}}\leq \frac{1}{3}$, it does not increase linearly like $g(\delta_E)=\delta_E$.\\
\section{Performance of EMICoRe vs. PEDT}
\indent In this section, we investigate the performance of EMICoRe and PEDT for 10 qubit Ising Hamiltonians, both off and at criticality. All trials were independently seeded and ran for 310 iterations.\\
\begin{figure}[H]
    \centering
    \begin{subfigure}[b]{0.45\textwidth}
    \includegraphics[width=\linewidth]{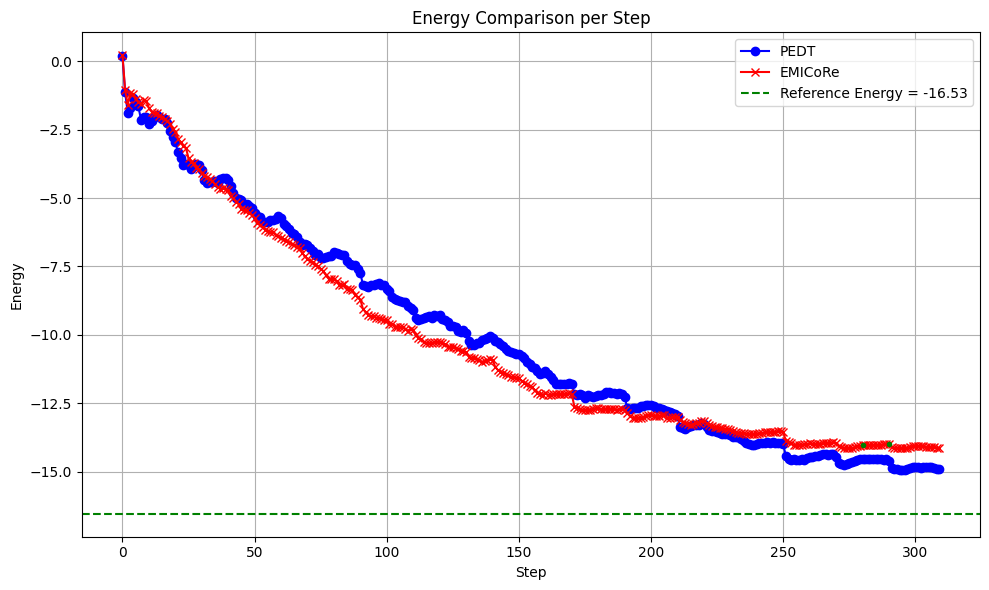}
    \caption{}
    \label{fig:Figure 3}
    \end{subfigure}
    \hfill
    \begin{subfigure}[b]{0.45\textwidth}
    \centering
    \includegraphics[width=\linewidth]{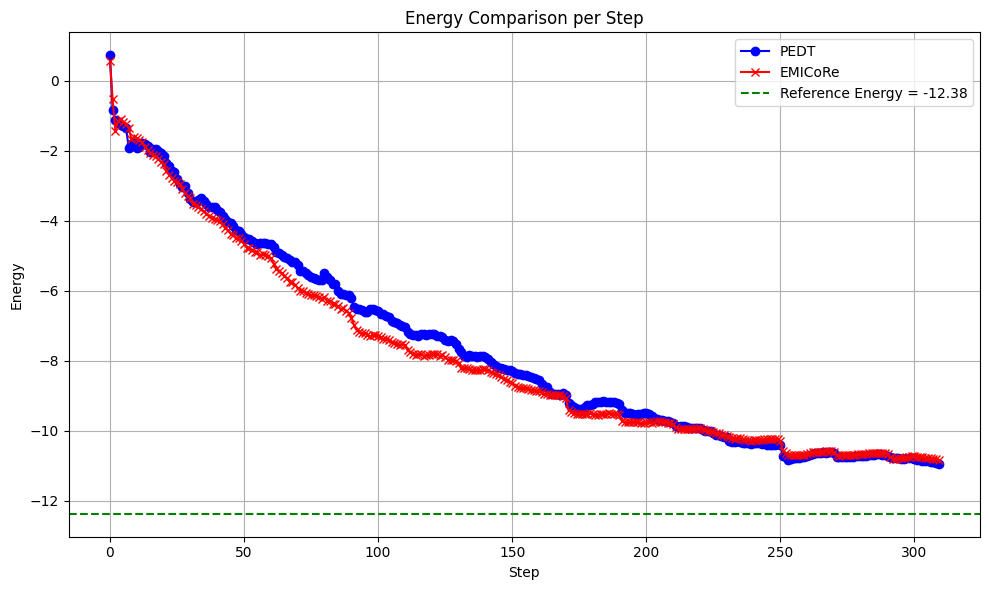}
    \caption{}
    \label{Figure 4}
    \end{subfigure}
    \caption{The 10 qubit Ising Hamiltonian off-criticality (a) and at criticality (b) ran for 310 iterations and 10 independent seeds. The average results over the 10 seeds are shown in blue for PEDT and red for EMICoRe and 2 points for which $\kappa=0$ between them are shown in green. The analytically computed ground state energy is shown through the green, dotted line.}
    \label{fig:figs3and4}
\end{figure}
\indent For the Ising Hamiltonian off-criticality we can see that our intuitions were correct. PEDT performed slightly better than the EMICoRe model in the initial stages of optimization and in the later steps PEDT continued to decrease and outperform EMICoRe, ultimately getting closer to the true ground state energy than the EMICoRe model. We can also experimentally see several instances where $\kappa$ becomes 0 for EMICoRe. One such example can be seen in iterations 280 and 290, where the EMICoRe model returned an energy value of -14.023 for iteration 280 but for iteration 290 the model returned a value of -13.999; see the green points in Fig. \ref{fig:Figure 3}. Using the EMICoRe group's definition of the threshold and $T_{Avg}=10$ as it was in all trials \cite{emicorecode}, $\kappa=$max$(0,\frac{\mu^{t-T_{Avg}}-\mu^t}{T_{Avg}})=0$ meaning the Confident Region can only be points with predicted 0 variance, or directly observed points, for the next 10 trials, as there is observation noise so every point predicted by the GP has some variance associated with it. This does not happen for the PEDT threshold as $\kappa_{PEDT}\neq0$, increasing the convergence rate in the final stages of optimization compared to EMICoRe, as can be seen in Fig. \ref{fig:Figure 3}.\\
\indent  The at criticality Ising Hamiltonian results shown in Fig. \ref{Figure 4} highlight the generality of the PEDT model's usefulness as it still performs similarly to the EMICoRe model, ending with an energy slightly closer to the ground states than EMICoRe's, but not significant enough to warrant claim of an improvement. The more difficult optimization regime presented by the at criticality system had a significant effect on the improvements made in the off-criticality regime but they are still present. This can be seen in Fig. \ref{Figure 4} where at iteration 270 the two models have near identical values but by iteration 310 PEDT has reached a lower energy than EMICoRe, largely due to EMICoRe's choice of an exceedingly harsh threshold.\\

\begin{figure}[H]
    \centering
    \includegraphics[width=0.5\linewidth]{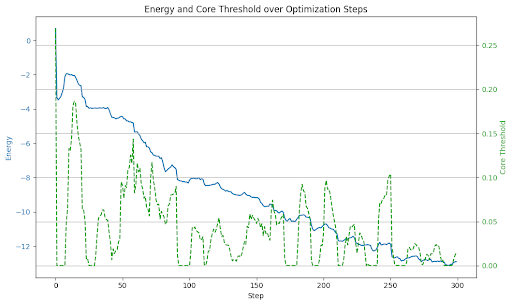}
    \caption{Evolution of the EMICoRe threshold and energy values as optimization occurs for the 10 qubit Ising Hamiltonian off-criticality for 300 iterations.}
    \label{fig:Figure 1}
\end{figure}
\indent Fig. \ref{fig:Figure 1} demonstrates how the EMICoRe threshold value affects energy convergence. We graph the evolution of the EMICoRe threshold simultaneously with the predicted energy to better understand the effect of the threshold on convergence. A key feature in Fig. \ref{fig:Figure 1} is the instances where $\kappa=0$. We discussed earlier that with the EMICoRe group's choice of threshold, $\kappa$ would return 0 at the beginning and final stages of optimization. This is supported by Fig. \ref{fig:Figure 1} as $\kappa=0$ at multiple occasions in the initial iterations (iterations 0-50) and the concluding iterations (iterations 250-300). With the energy at each iteration overlayed on the threshold graph, we can see that $\kappa$ being 0 negatively affects the rate of convergence, often prompting slower decreases or even increases in predicted energy for the following iterations.\\
\indent As a final test we generalize the Ising system to the Heisenberg Hamiltonian, a common benchmark with VQEs due to its applications in condensed matter physics, specifically with magnetic materials \cite{mrwhite}. We compare performance between the two models for the 10 qubit Heisenberg Hamiltonian ran for 310 iterations.\\
\begin{figure}[H]
    \centering
    \includegraphics[width=0.5\linewidth]{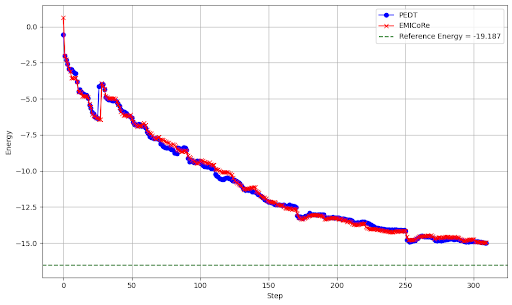}
    \caption{The 10 qubit Heisenberg Hamiltonian ran for 310 iterations and 10 independent seeds. The average results over the 10 seeds are shown in blue for PEDT and red for EMICoRe. The analytically computed ground state energy is shown through the green, dotted line.}
    \label{fig:Figure 5}
\end{figure}

\indent For the Heisenberg system, performance between the two systems is practically identical, with both models reaching similar ending energies after optimization. Both the Ising Hamiltonian at criticality and the Heisenberg Hamiltonian represent rougher energy landscapes where more local minima are present, making optimization more difficult for VQEs. For example, a spin glass system consisting of a random 2D Ising model has an approximately Gaussian distribution of local minima, making it easier for optimization to become trapped in suboptimal local minima \cite{spinglass}. However, utilizing these systems and similar ones to them is more indicative of capability for practical applications of the model. The enhanced presence of these local minima hinder performance, especially for less explorative models, forcing a precise exploration-exploitation balance \cite{master}. The introduction of the PEDT threshold with its chosen numerical parameters is a more explorative choice of threshold when compared to EMICoRe but still resulted in similar or slightly better performance between the two models for the more difficult systems. However, the framework provided by the PEDT threshold has several unoptimized degrees of freedom and the results presented in this paper suggest that further research into these has the potential to yield further advancements in VQE performance.
\section{Conclusion}
\indent Although the intuition behind the PEDT model proved to be correct and the results suggest improvements or similar performance over EMICoRe in several key optimization regimes, there are several areas to be explored that could lead to further improvements, even for more complex systems to be approximated. The choice of PEDT, Equation~\eqref{eq:1}, has predetermined  numerical constants. These chosen constants performed better than constants that produce more lenient or harsher thresholds; see Appendix A. However, a more general framework for PEDT looks like $$\kappa_{PEDT}=(a\cdot \sigma_0)\cdot \text{min}(\delta_E,\frac{b}{c+d\cdot e^{-f\cdot \delta_E}})$$ where $a$ through $f$ are constants to optimize in further studies. This new choice for PEDT or even a more sophisticated threshold would likely perform better than the current model and is left to further investigation. Additionally, measuring performance for larger systems not restricted to Ising or Heisenberg Hamiltonians for both the EMICoRe and the PEDT models 
would better test their robustness and performance outside of the Ising system.\\
\indent We have shown that by utilizing a more forgiving, robust threshold and eliminating instances where $\kappa=0$,  performance can be on par with or even surpass EMICoRe. By incorporating information from the GP prior and the change in energy, a lower convergence value can be reached for the Ising Hamiltonian both at and off criticality. \\
\indent Continuing to improve on VQE algorithms has the potential for developments in quantum chemistry and materials sciences as it is one of the few quantum methods that is practically implementable in research due to blending of classical and quantum elements. This blending leads to less reliance on inherently uncertain quantum system elements while still maintaining its superior performance over classical counterparts, an advantage that is pivotal to advances in these fields and is worth progressing.
\newpage
\printbibliography
\newpage
\section*{Appendix A}
\begin{figure}[h]
    \centering
    \includegraphics[width=0.5\linewidth]{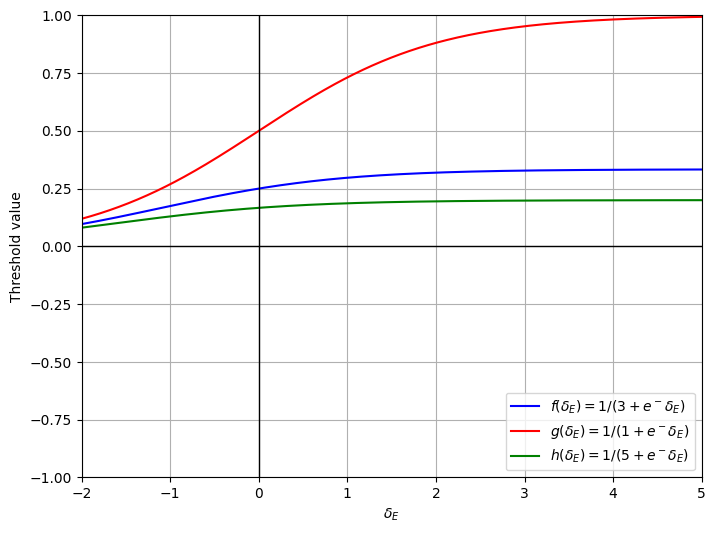}
    \caption{Graphs of a more lenient and strict threshold compared to the version employed in the main paper.}
    \label{Figure 6}
\end{figure}

\begin{figure}[H]
    \centering
    \includegraphics[width=0.5\linewidth]{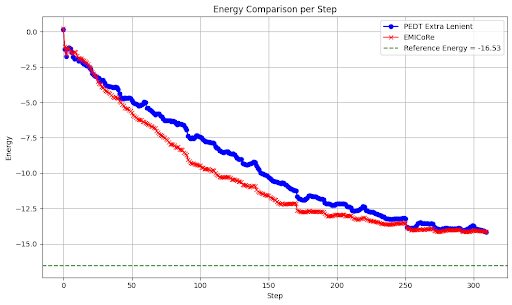}
    \caption{The 10 qubit Ising Hamiltonian off-criticality ran for 310 iterations and 10 independent seeds. The average results over the 10 seeds are shown in blue for PEDT Extra Lenient and red for EMICoRe. The analytically computed ground state energy is shown through the green, dotted line}
    \label{Figure 7}
\end{figure}

In the appendix we analyze more lenient and strict choices for PEDT to justify the choice employed in the comparisons against EMICoRe. We restrict our analysis to $c$ in the more generalized framework of $\frac{b}{c+d\cdot e^{-f\cdot \delta_E}}$ and leave the investigation of the other constants to further studies. We test the EMICoRe threshold versus PEDT with the $\kappa_1=\text{min}(\delta_E,\frac{1}{1+e^{-\delta_E}})$ and $\kappa_2=\text{min}(\delta_E,\frac{1}{5+e^{-\delta_E}})$ with $\kappa_1$ being a more lenient threshold and $\kappa_2$ being a stricter threshold compared to the one used in the paper, see Fig. \ref{Figure 6}.
\\
We test the effectiveness of these thresholds in the 10 qubit Ising Hamiltonian off-criticality regime and optimization is run for 310 iterations for 10 independently seeded trials. Run with $\kappa_1$ the model returns Fig. \ref{Figure 7}:\\
The extra lenient threshold is more volatile in the later stages of optimization compared to the version of PEDT used in the main paper (compare to Fig. \ref{fig:Figure 3}) and instead of consistent decreases, the energy is shown to increase at several instances, highlighting the ramifications of a very inclusive threshold. Having such an inclusive region leads to the introduction of low quality (high variance) points into the CR which negatively affects optimization.\\
Similarly, for $\kappa_2$ the model returns Fig. \ref{Figure 8}:
\begin{figure}[H]
    \centering
    \includegraphics[width=0.5\linewidth]{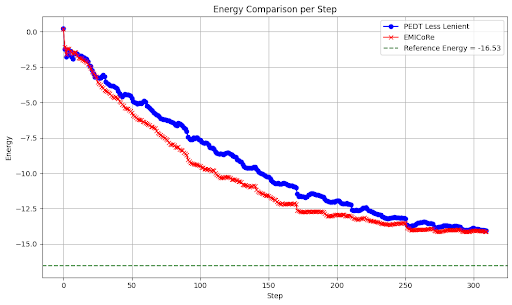}
    \caption{The 10 qubit Ising Hamiltonian off-criticality ran for 310 iterations and 10 independent seeds. The average results over the 10 seeds are shown in blue for PEDT Less Lenient and red for EMICoRe. The analytically computed ground state energy is shown through the green, dotted line}
    \label{Figure 8}
\end{figure}
Choosing an excessively restrictive threshold allows very few points into the CR. This leads to the model having very few points to choose from when optimizing for the following iteration. While the points contained in the CR have very little variance, confining the CR to such a small set of points is contradictory to its proposed purpose, which was to broaden the set of points the model can test to use in the ensuing rounds of optimization. This can be seen especially in the middle stages of optimization where a moderately sized CR is desired. Fig. \ref{Figure 8} displays this weakness as EMICoRe performs significantly better in the middle stages compared to PEDT Less Lenient (contrast with Fig. \ref{fig:Figure 2} where the difference between the two models is substantially less during the middle iterations).

\section*{Appendix B}
From \cite{gaussianprocess} we get an equation for the posterior covariance at a test point $(\sigma_*^2)$ in terms of the prior covariance between two test points $k(\mathbf{x}_*,\mathbf{x}_*)$: $$\sigma_*^2=k(\mathbf{x}_*,\mathbf{x}_*)-\mathbf{k}(\mathbf{x}_*)^{\intercal}(K(X,X)+\sigma_n^2I)^{-1}\mathbf{k}(\mathbf{x}_*)$$ where $\sigma_n^2$ is the Gaussian noise variance, $\mathbf{k}(\mathbf{x}_*)$ is the vector of covariances between a test point and $n$ training points, $\intercal$ is the transpose, $K(X,X)$ is the matrix of covariances between training points, and $I$ is the identity matrix of dimension equal to $K(X,X)$ . The second term is a quadratic form associated with a covariance matrix and is thus positive semi-definite. This means that the second term is greater than or equal to $0$ and thus  $\sigma_*^2\leq  k(\mathbf{x}_*,\mathbf{x}_*)$ which justifies the use of the prior variance as an upper bound. For a more thorough treatment of GPs see \cite{gaussianprocess}.

\end{document}